\documentclass[aps,showpacs,prl,twocolumn,superscriptaddress]{revtex4}

\usepackage{graphicx}
\usepackage{dcolumn}
\usepackage{bm}
\usepackage{epsfig}

\newcommand{\bOmega}{{\bm{\Omega}}}
\newcommand{\rmd}{\textrm{d}}
\newcommand{\here}{\makebox(0,0)}
\newcommand{\R}{{\rm I\!R}}

\begin{document}

\preprint{FOG_2005_1}

\title{Dynamical replica analysis of disordered Ising spin systems on 
finitely connected random graphs}

\author{J.P.L. Hatchett}
\affiliation{Laboratory for Mathematical Neuroscience, RIKEN Brain
  Science Institute, Wako, Saitama 351-0198, Japan}
\email{hatchett@brain.riken.jp}

\author{I. \surname{P\'erez Castillo}}
\affiliation{Rudolph Peirls Center for Theoretical Physics, University
  of Oxford, 1 Keble Road, Oxford OX1 3NP, UK}
\email{isaac@thphys.ox.ac.uk}

\author{A.C.C. Coolen}
\affiliation{Department of Mathematics, King's College, University of
  London, The Strand, London WC2R 2LS, UK}
\email{tcoolen@mth.kcl.ac.uk}

\author{N.S. Skantzos}
\affiliation{Instituut voor Theoretische Fysica,
  Katholieke Universiteit Leuven, Celestijnenlaan 200, DB-3001 Leuven, Belgium}
\email{nikos@itf.fys.kuleuven.ac.be}

\date{\today}

\begin{abstract}
We study the dynamics of macroscopic observables such as the
magnetization and the energy per degree of freedom in Ising spin
models on random graphs of finite connectivity, with random bonds
and/or heterogeneous degree distributions. To do so we generalize
existing implementations of dynamical replica theory and cavity
field techniques to systems with strongly disordered and locally
tree-like interactions. We illustrate our results via application
to the dynamics of e.g.~$\pm J$ spin-glasses on random graphs and
of the overlap in finite connectivity Sourlas codes.
All results are tested against Monte Carlo simulations.
\end{abstract}

\maketitle

Recent years have witnessed a surge of interest in the study of
finitely connected  disordered spin systems. From a physical point
of view, despite their lack of a realistic geometry and their
mean-field nature, the finite number of neighbours per spin in
such models does give rise to a non-trivial local geometry and
ensuing artifacts. Here we simply regard the random bond finite
connectivity Ising spin system as the archetypal interacting
particle model on a disordered random graph. Such models are
important in the understanding of algorithmic complexity in
theoretical computer science
\cite{Mezard2002,Monasson1999b,Martin2001}, and also underly
recent theoretical advances for an important class of error
correcting codes \cite{Kabashima1999a,Kanter2000a,Kabashima2004}.
It has been shown that the tuning of the degree distribution and/or
the connectivity strengths in complex networks can lead to
atypical mean field critical phenomena
\cite{Leone2002,Goltsev2002,Giuraniuc2004}, and they are now
increasingly and fruitfully used for modeling neural, social, internet,
gene regulatory
and proteomic networks
\cite{Wemmenhove2003,Castillo2004,Galam1982,Correale2004,Dorogovtsev2001}.
While our quantitative understanding of the equilibrium properties
of these systems is quite advanced (see
e.g.~\cite{Mezard2001,Mezard2003a}), the tools available for studying 
their non-equilibrium behaviour are comparatively poor. There has
been some progress in applying the path integral techniques of
\cite{DeDominicis1978} to spherical and related models
\cite{Semerjian2003a}, and to Ising models with parallel spin
updating \cite{Hatchett2004,Hatchett2005}. Generalizing such
approaches to Ising spin models with Glauber-type dynamics
requires the treatment of non-trivial functional order parameters
which have, as yet, not been adequately controlled. An alternative
approach, which we follow here, is to generalize the techniques of
dynamical replica theory (DRT) \cite{Coolen1994,Laughton1996},
together with the cavity field concept, to finitely connected
disordered spin systems. This approach has already proven fruitful
for weakly disordered dilute ferromagnets \cite{Semerjian2004b}
where each spin is effectively in an identical environment. In
this letter, in contrast, we study the dynamics of strongly
disordered versions of finitely connected Ising systems, where
each spin is in a highly {\em heterogeneous} environment, due to
the presence of either random bonds or nodes with variable
degrees.

Our model consists of $N$ Ising spins $s_i \in \{-1,1\},~ i =
1,\ldots, N$, whose mutual interactions are characterized by a
range-free symmetric adjacency matrix with entries
$c_{ij}\in\{0,1\}$ and symmetric bonds $J_{ij}\in\R$. We define
$c_{ii} = 0$, and draw the bond strengths $J_{ij}$ i.i.d.~from
some distribution $Q(J)$. The probability of finding any state
$\mathbf{s} \equiv (s_1,\ldots,s_N)$ of the system at time $t$ is
given by the measure $p_t(\mathbf{s})$, which evolves as the
spins align asynchronously and stochastically to their local
fields, according to a Glauber dynamics in the form of the master
equation
\begin{equation}
\frac{\rmd}{\rmd t}p_t(\mathbf{s}) = \sum_{k = 1}^N[p_t(F_k
  \mathbf{s}) w_k(F_k\mathbf{s}) - p_t(\mathbf{s}) w_k(\mathbf{s})]
\end{equation}
where $F_k\mathbf{s} \equiv (s_1,\ldots,-s_k,\ldots,s_N)$ is the
$k$th spin-flip operator and the transition rates
$w_k(\mathbf{s})$ have the standard form
\begin{equation}
w_k(\mathbf{s}) \equiv \frac{1}{2}\left\{1 - s_k \tanh[\beta
  h_k(\mathbf{s})]\right\}
\end{equation}
with the local fields $h_i(\mathbf{s}) \equiv \sum_{j\neq i}
c_{ij} J_{ij} s_j + \theta$. This process evolves toward
equilibrium, with the Boltzmann measure and with Hamiltonian
\begin{equation}
H(\mathbf{s}) = -\sum_{i < j} s_i c_{ij} J_{ij} s_j - \theta \sum_i s_i
\label{eq:hamiltonian}
\end{equation}

Following the procedure outlined for fully connected systems
\cite{Coolen1994,Laughton1996} we consider the evolution of two
macroscopic observables, the magnetization $m(\mathbf{s}) = N^{-1}
\sum_i s_i$ and the internal energy $e(\mathbf{s}) = N^{-1}
\sum_{i < j} c_{ij} J_{ij} s_i s_j$. We will abbreviate
$\bOmega=(m,e)$. One easily derives a Kramers-Moyal (KM) expansion
for their probability density
$P_t(\bOmega)=\sum_{\mathbf{s}}p_t(\mathbf{s})
\delta[\bOmega-\bOmega(\mathbf{s})]$. On finite times one finds
that only the Liouville term in the KM expansion survives the
thermodynamic limit, so the observables $(m,e)$ evolve
deterministically, i.e.~$\lim_{N\to\infty}P_t(\bOmega)=\delta[\bOmega-\bOmega_t]$ with
\begin{eqnarray}
\frac{\rmd}{\rmd t} \bOmega_t = 
\lim_{N\to\infty}\sum_{\mathbf{s}}p_t(\mathbf{s}|\bOmega_t)
\sum_i w_i(\mathbf{s})[\bOmega(F_i\mathbf{s})-\bOmega(\mathbf{s})]
\label{eq:dOmegadt}
\\
p_t(\mathbf{s}|\bOmega)= \frac{ p_t(\mathbf{s}) \delta[\bOmega -
\bOmega(\mathbf{s})] }{\sum_{\mathbf{s}^\prime}
p_t(\mathbf{s}^\prime) \delta[\bOmega -
\bOmega(\mathbf{s}^\prime)]}\hspace*{20mm}
\end{eqnarray}
Equation (\ref{eq:dOmegadt}) still involves the conditional
microscopic distribution $p_t(\mathbf{s}|\bOmega)$. To proceed we
follow \cite{Coolen1994,Laughton1996}: we (i) assume that the
observables $\bOmega$ are self-averaging at all times (which one
expects to be true), and (ii) approximate the microscopic measure
$p_t(\mathbf{s}|\bOmega)$ by the maximum entropy distribution
given the condition that the macroscopic observables take the
value  $\bOmega$. These assumptions imply that our observables
evolve according to
\begin{eqnarray}
\frac{\rmd}{\rmd t} m_t = -m_t + \int\! \rmd h~ \mathcal{D}(h|m_t,
e_t) \tanh(\beta h) \label{eq:magnetisation_evolution}\\
\frac{\rmd}{\rmd t} e_t = -2e_t - \int\! \rmd h~
\mathcal{D}(h|m_t,e_t) h \tanh(\beta h)
\label{eq:energy_evolution}
\end{eqnarray}
Here $\mathcal{D}(h|m_t,e_t)$ denotes the asymptotic distribution
of local fields in a system with magnetization $m_t$ and energy
$e_t$,
\begin{equation}
\mathcal{D}(h|e_t,m_t) =\lim_{N\to\infty}  \frac{1}{N} \sum_{i=1}^N \left[
\left\langle \delta[h -
  h_k(\mathbf{s})] \right\rangle_{m_t, e_t} \right]_{dis}
\end{equation}
where the average $\left\langle \ldots \right\rangle_{m_t,e_t}$ is
over the maximum entropy distribution given the values of the
observables, viz. over
\begin{equation}
p(\mathbf{s}|m_t,e_t)\equiv \frac{\delta[m_t -m(\mathbf{s})]
\delta[e_t - e(\mathbf{s})] }{\sum_{\mathbf{s}^\prime} \delta[m_t
- m(\mathbf{s}^\prime)]\delta[e_t - e(\mathbf{s}^\prime)]}
\end{equation}
and $[\ldots]_{dis}$ is over the disorder
(the realization of the random graph and bonds).

The relatively
simple solution given in \cite{Semerjian2004b} can be understood
within the current framework. Since all sites were identical,
there was just a single cavity field and hence the distribution of
local fields was uniform across sites and could be given
explicitly in terms of the observables.

The field distribution $\mathcal{D}(h|m_t,e_t)$ is
readily calculated, even in the presence of bond or degree disorder,
either via the replica
method or via the cavity approach for dilute systems
\cite{Mezard2001} (in the microcanonical or the canonical
framework, respectively)\footnote{Although the replica approach is
more general since it does not require the existence of a
Hamiltonian, i.e.~it allows also for the study of systems without
detailed balance.}. Here the resulting equations from either
approach are the same. They correspond to the maximum entropy
distribution, given $(m_t,e_t)$, which equals the equilibrium
distribution of a system with Hamiltonian (\ref{eq:hamiltonian})
but with a pseudo inverse temperature $\hat{\beta}$ and pseudo
external field $\hat{\theta}/\hat{\beta}$. The latter act as Lagrange
parameters, enforcing the condition that the equilibrium
distribution gives the required values of $(m_t, e_t)$. The prefix
`pseudo' indicates that these parameters need not be physical:
there could be states $(m_t,e_t)$ for which $\hat{\beta}$ is
negative. Within the cavity formalism we can work either with the
ensemble or with a particular realization of the disorder. The
latter tends to be numerically simpler, due to the inherent
(finite size) noise in population dynamics in the ensemble, which
limits the accuracy with which the Lagrange parameters can be
calculated.  Working in areas of phase space where replica
symmetry is expected to be exact and where belief propagation
converges on any given graph realization, it is possible to find
the Lagrange parameters to high precision. For large graph sizes
the differences between results for graph realization and the
ensemble average ought to vanish.

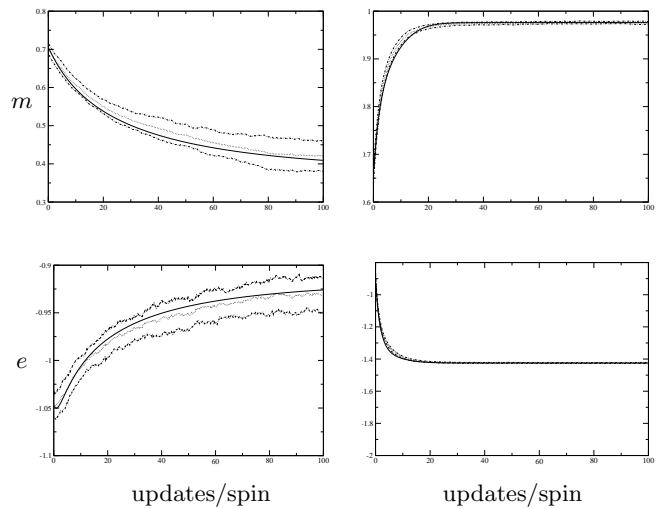
\begin{figure}[t]
\vspace*{5mm} \hspace*{5mm} \setlength{\unitlength}{0.45mm}
\begin{picture}(180,140)
\put(90,90){\epsfysize=60\unitlength\epsfxsize=87\unitlength\epsfbox{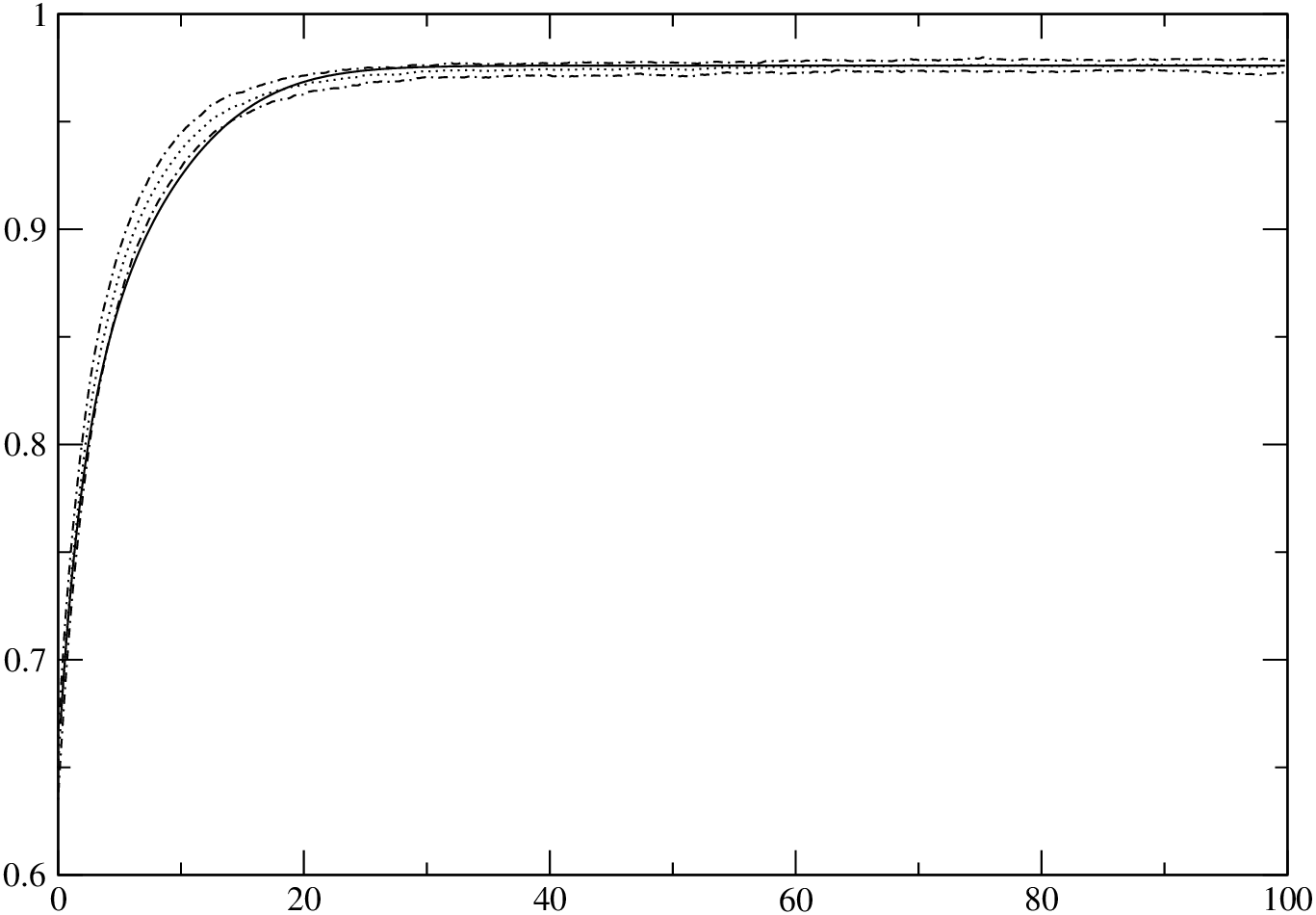}}
\put(-6,90){\epsfysize=60\unitlength\epsfxsize=87\unitlength\epsfbox{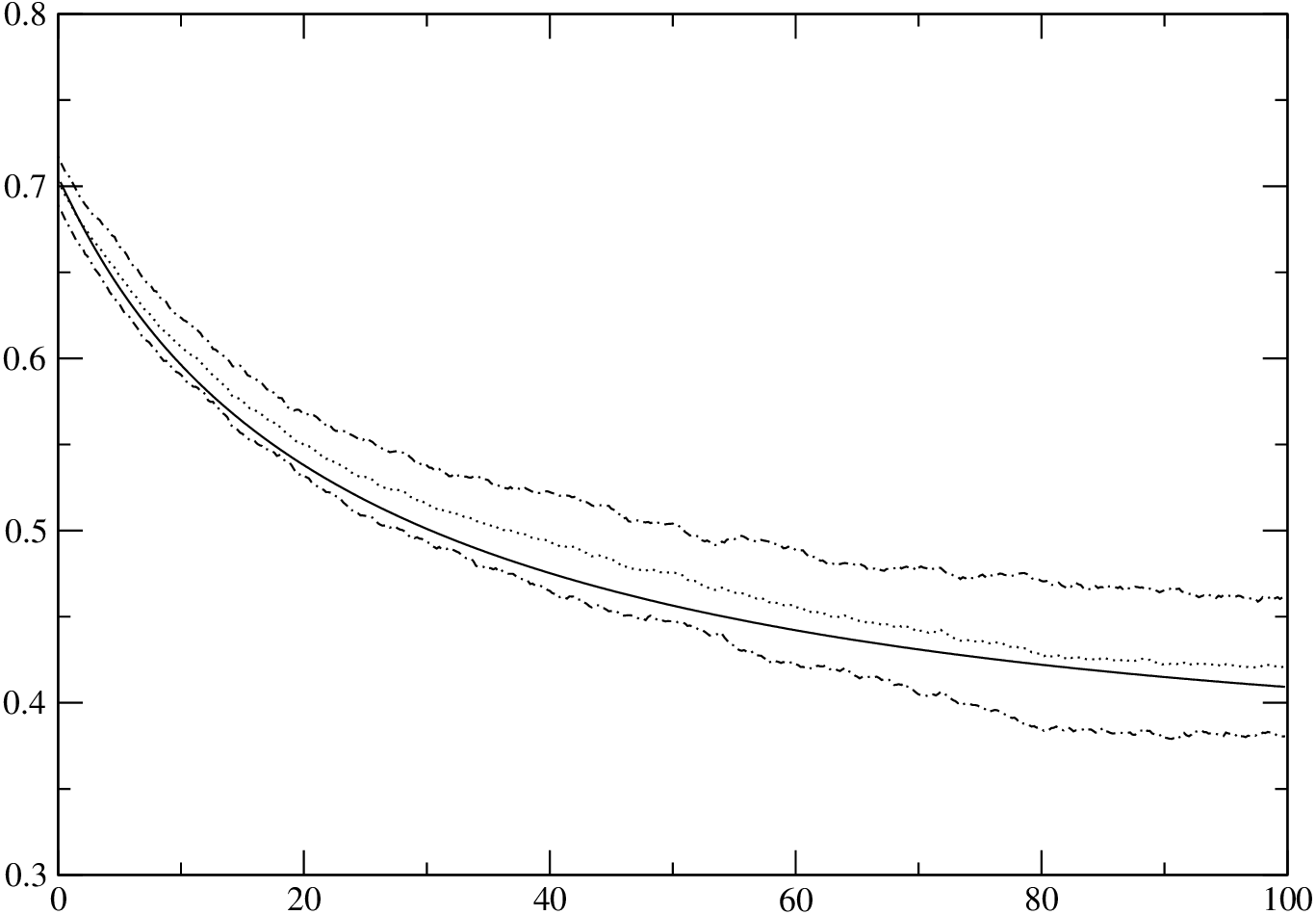}}
\put(90,15){\epsfysize=60\unitlength\epsfxsize=87\unitlength\epsfbox{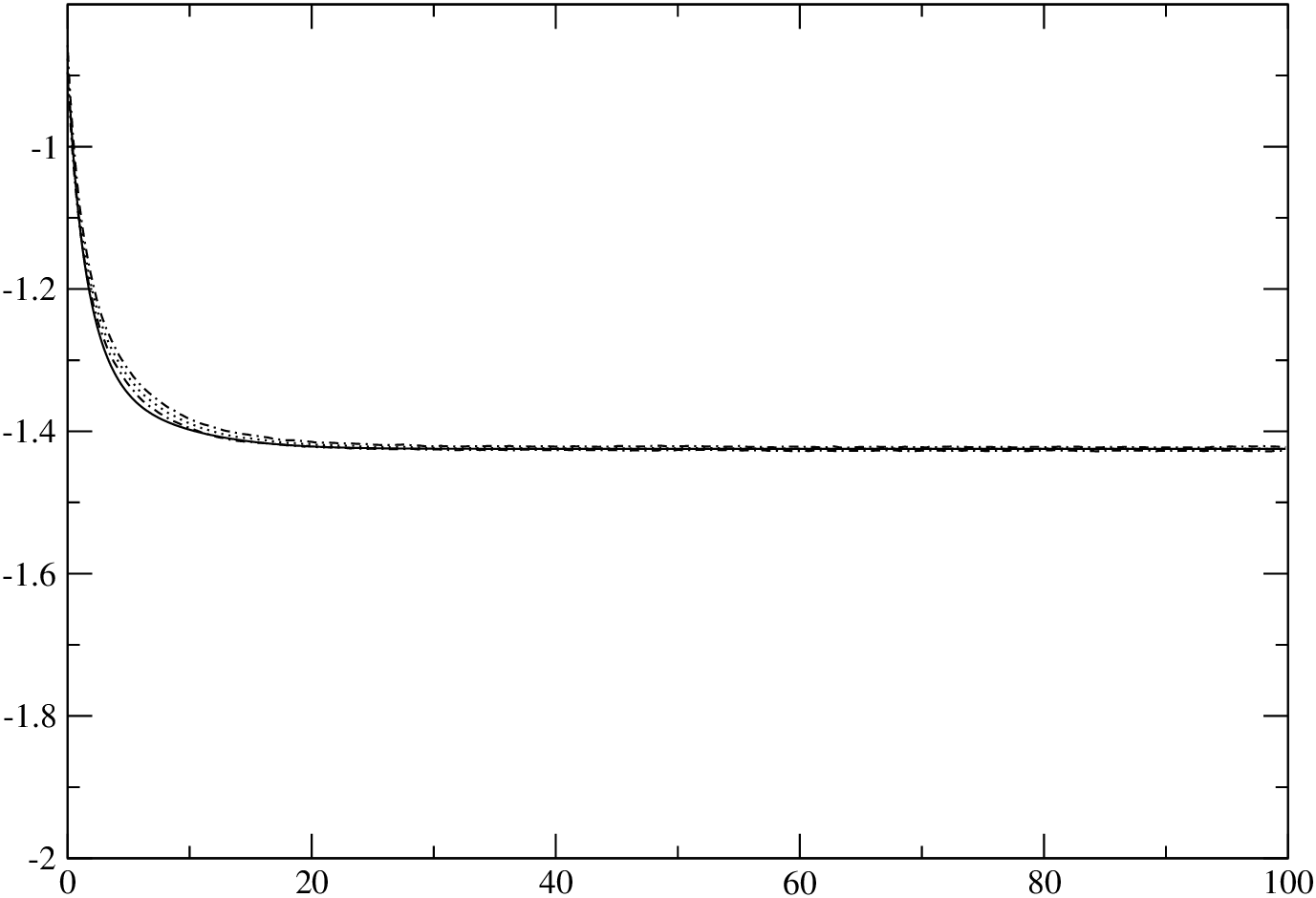}}
\put(-6,15){\epsfysize=60\unitlength\epsfxsize=87\unitlength\epsfbox{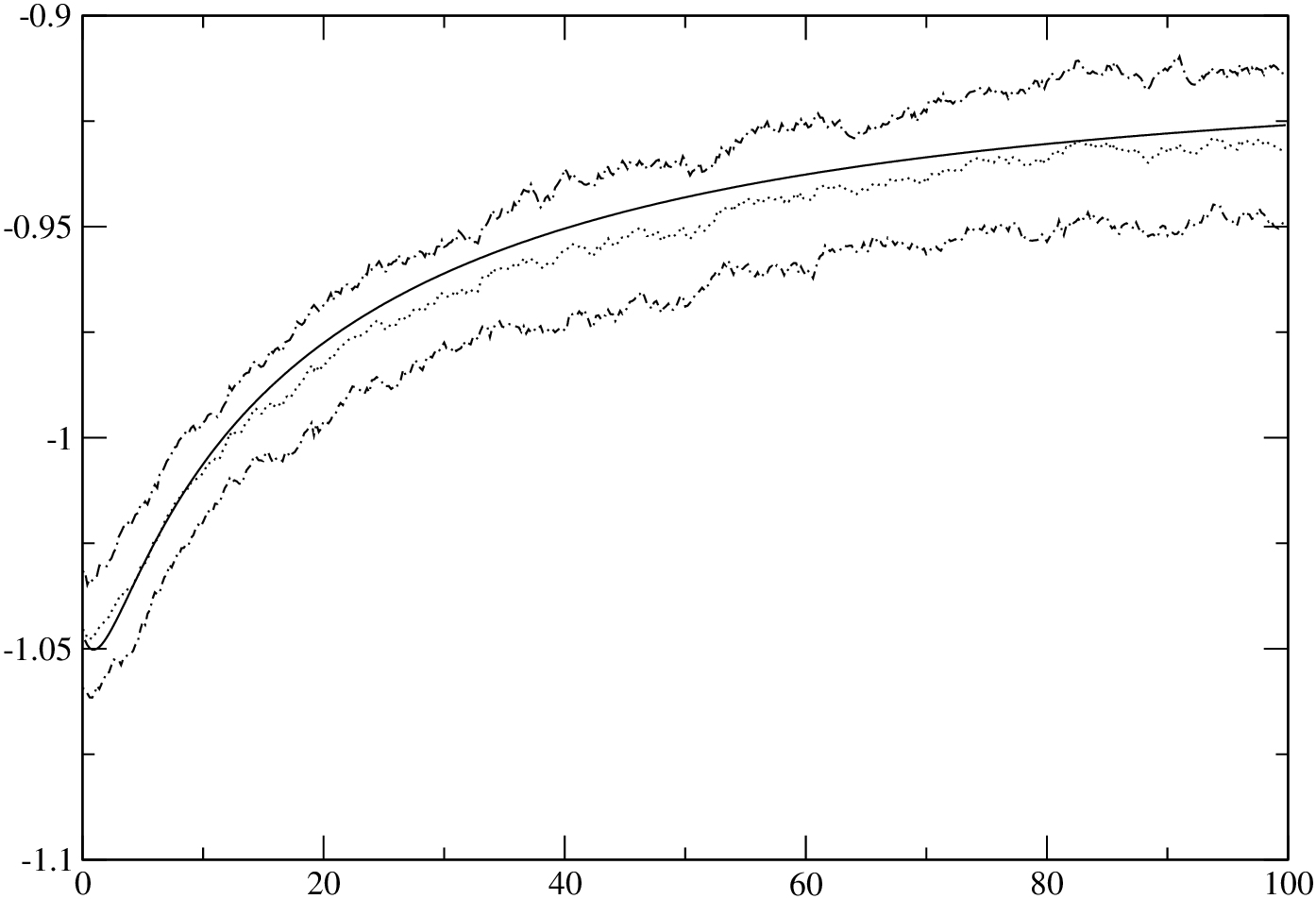}}
\put(43,6){\here{\small updates/spin}}
\put(135,6){\here{\small updates/spin}}
\put(-10,45){\here{$e$}}
\put(-10,122){\here{$m$}}
\end{picture}
\vspace*{-5mm} \caption{\label{fig:sgevol} Evolution of the
magnetization $m$ (top figures) and the energy $e$ (bottom
figures), for Ising spins on a 3-regular random graph with random
bonds, and with time measured in units of updates per spin. Bond
distribution:  $Q(J) = \eta \delta(J-1) + (1-\eta) \delta(J+1)$.
Solid lines denote the theoretical predictions. Dotted lines
represent the simulation data (system size $N=10,\!000$ and
averaged over 50 runs), with dot-dashed lines giving the averages
$\pm$ 1 standard deviation. Left pictures: $\eta = 0.95$ and
$\beta = 0.65$. Right pictures: $\eta = 0.97$ and $\beta =
1.2$.\vspace*{-6mm} }
\end{figure}

The resulting numerical algorithm is as follows. At any given
point in time we know the instantaneous values ($m_t,e_t$) of our
observables. We then run a belief propagation algorithm on our
graph, for a given pair of Lagrange parameters ($\hat{\beta},
\hat{\theta}$) which act as inverse temperature and external
field; once the belief propagation has converged we can measure
($m_t(\hat{\beta},\hat{\theta}), e_t(\hat{\beta},\hat{\theta})$).
We now vary the Lagrange parameters and repeat the above until we
satisfy the condition
\begin{eqnarray}
m_t = m_t(\hat{\beta},\hat{\theta}),\quad e_t =
e_t(\hat{\beta},\hat{\theta})
\label{eq:conditions}
\end{eqnarray}
Finally we use the cavity fields generated with the correct values
of $(\hat{\beta}, \hat{\theta})$ to give the local field
distribution within our graph, with which we can evaluate the
force terms in
(\ref{eq:magnetisation_evolution},\ref{eq:energy_evolution}).

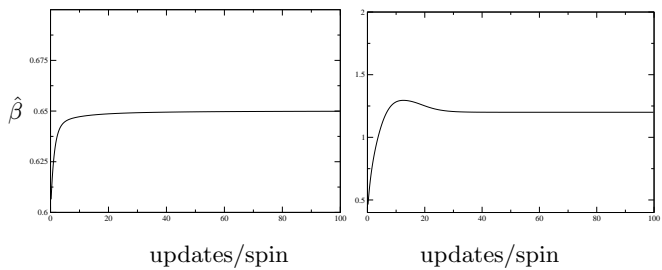
\begin{figure}[t]
\vspace*{5mm} \hspace*{5mm} \setlength{\unitlength}{0.45mm}
\begin{picture}(180,60)
\put(90,5){\epsfysize=63\unitlength\epsfbox{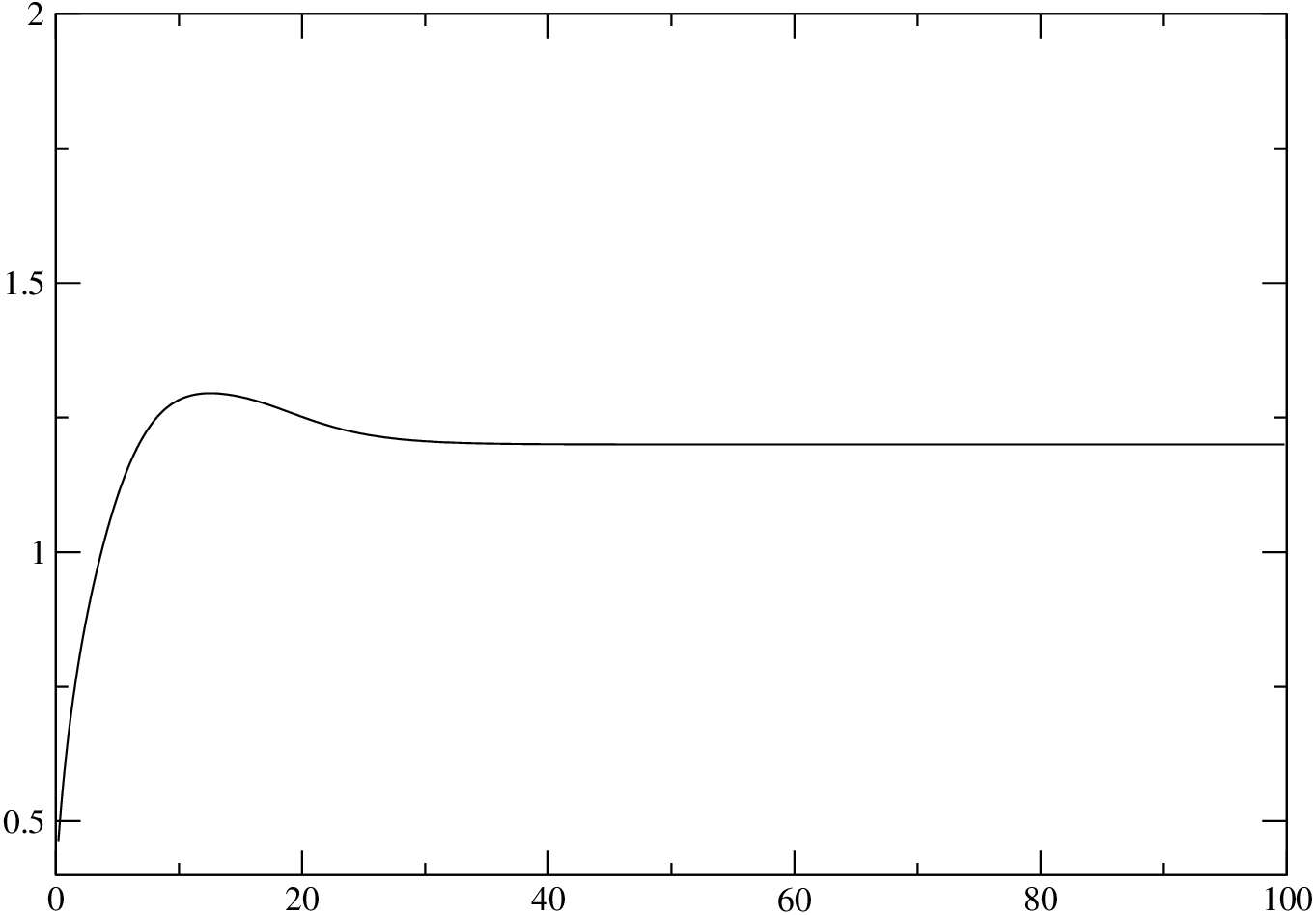}}
\put(-6,5){\epsfysize=63\unitlength\epsfbox{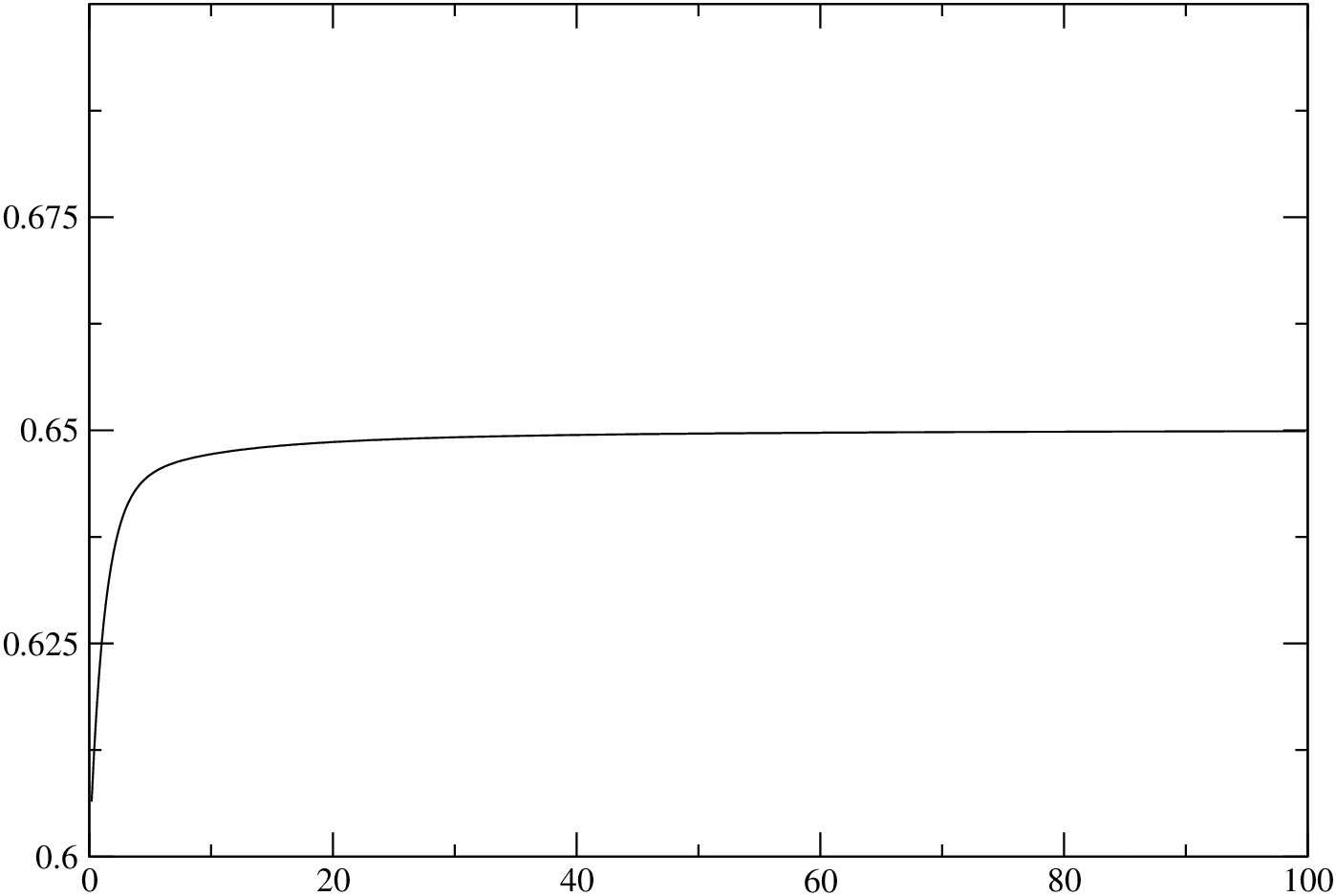}}
\put(-10,38){\here{\small $\hat{\beta}$}} 
\put(50,-5){\here{\small updates/spin}} 
\put(130,-5){\here{\small updates/spin}}
\end{picture}
\vspace{-2mm}\caption{\label{fig:beta_evol}
Evolution of the inverse pseudo-temperature Lagrange parameter
$\hat{\beta}$ for the experiments shown in figure
\ref{fig:sgevol}. Clearly, the relaxation of $\hat{\beta}$ need
not be monotonic.\vspace*{-6mm}}
\end{figure}

In figure \ref{fig:sgevol} we compare the results of our analysis
with Monte Carlo simulations for a $\pm J$ spin-glass on a
3-regular graph. We sampled our graph uniformly from all connected
graphs where each site has exactly three neighbours and each bond
is drawn i.i.d.~from $Q(J) = \eta \delta(J-1) + (1-\eta)
\delta(J+1)$. All simulations were carried out with a system size
of $N = 10,\!000$, and were run on the same realization of the
graph as the cavity field calculations. We see an excellent
correspondence between theory and simulations. We have taken
$\eta$ to be relatively large (predominance of ferromagnetic
bonds), since we did not wish to move into a region where belief
propagation would not converge, a condition one expects to be
closely related to instability in the AT sense
\cite{deAlmeida1978,Kabashima2003}. In such regions it is no
longer possible to use belief propagation for accurately evaluating
the Lagrange parameters $\hat{\beta}$ and $\hat{\theta}$.

It is also of interest to note in figure \ref{fig:beta_evol} that
the evolution of the pseudo-temperature need not be monotonic.
Assuming that the location of the AT line \cite{deAlmeida1978} is
similar to that of the fully connected case, i.e.~that it goes
continuously from the zero-temperature instability ($T = 0, \eta =
\frac{11}{12}$) \cite{Kwon1988,Castellani2004} to the triple point
($T \approx 1.13, \eta \approx 0.85$), as shown in
\cite{Kabashima2003}, one could envisage a situation where
starting from a replica symmetric (RS) phase, the parameters
($\beta, \eta$) could be chosen such that the final equilibrium
phase was also RS, but where the dynamics would take the system
through a regime of phase space where in equilibrium one would
find replica symmetry breaking. Since there, the belief propagation
(or any other replica symmetric) algorithm would not converge in a
time of $\mathcal{O}(N)$, the algorithm would become stuck {\em en
route} to RS equilibrium.

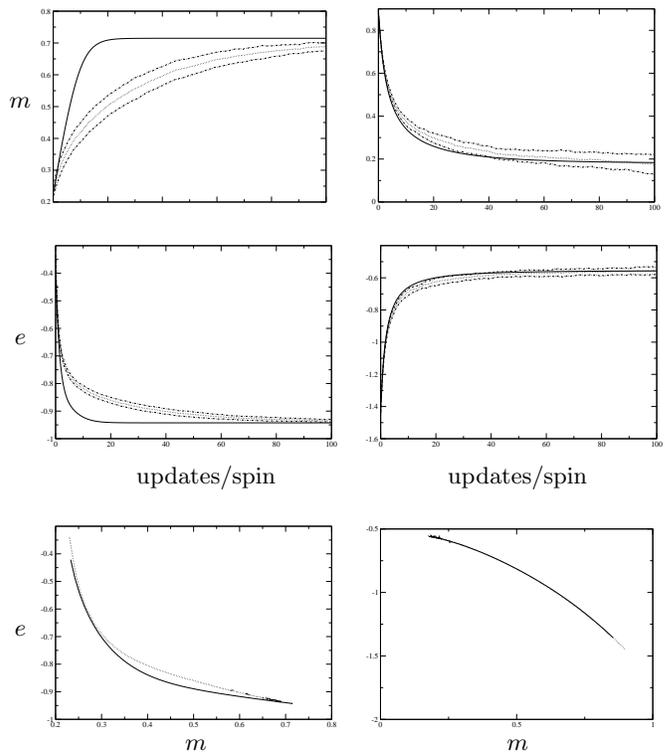
\begin{figure}[t]
\vspace*{5mm} \hspace*{5mm} \setlength{\unitlength}{0.45mm}
\begin{picture}(180,210)
\put(-6,7){\epsfysize=60\unitlength\epsfbox{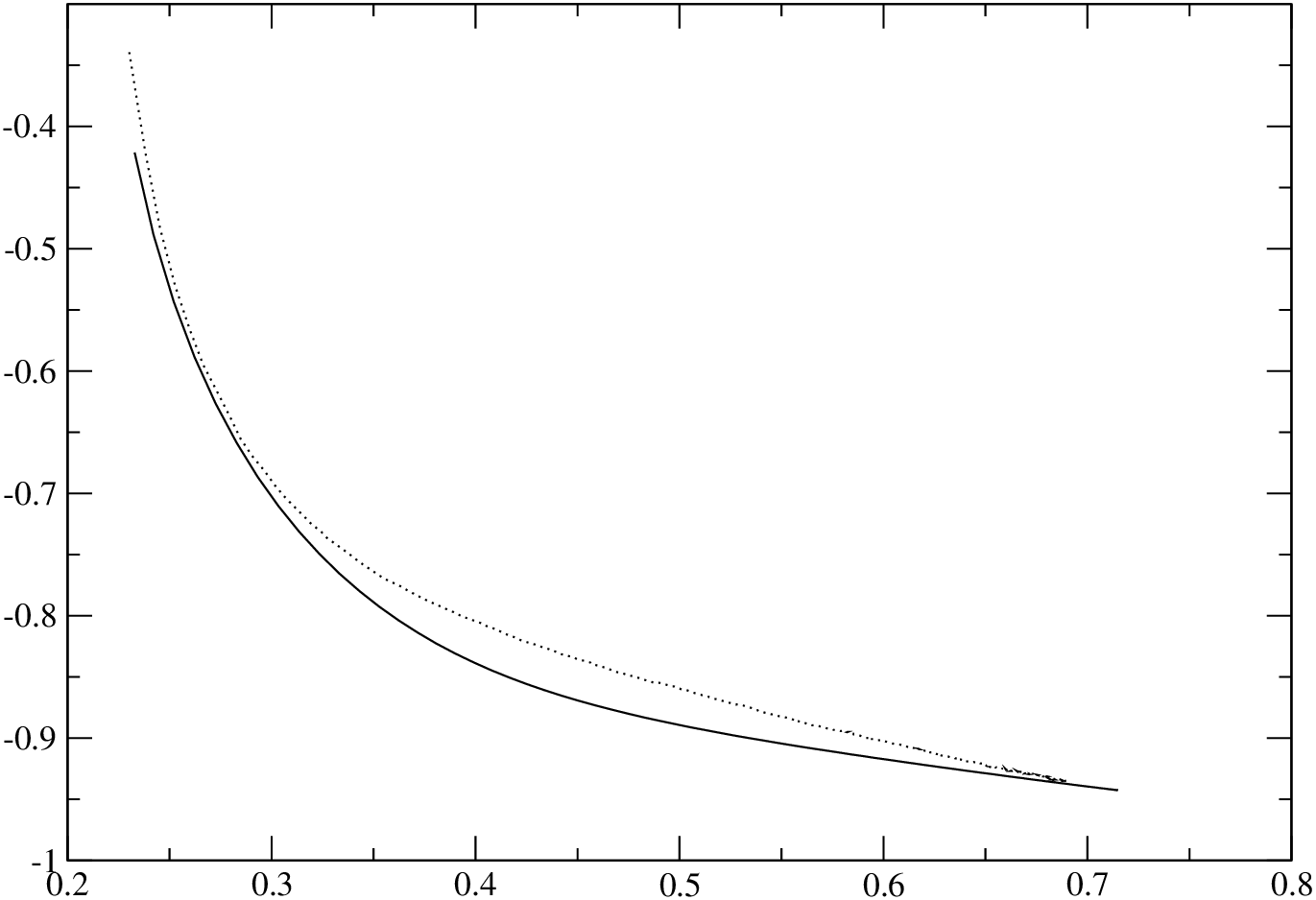}}
\put(90,7){\epsfysize=60\unitlength\epsfbox{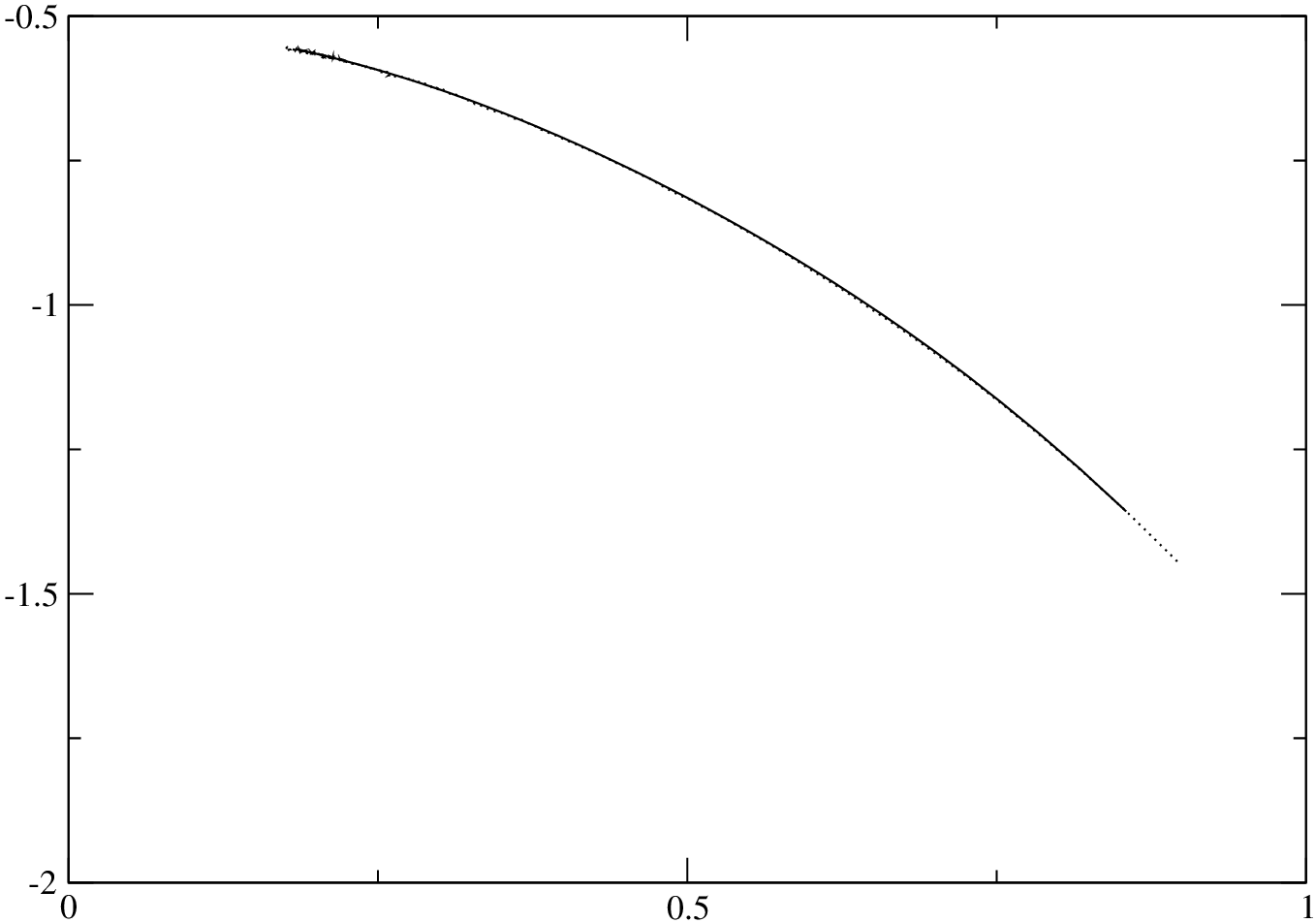}}
\put(-6,160){\epsfysize=60\unitlength\epsfbox{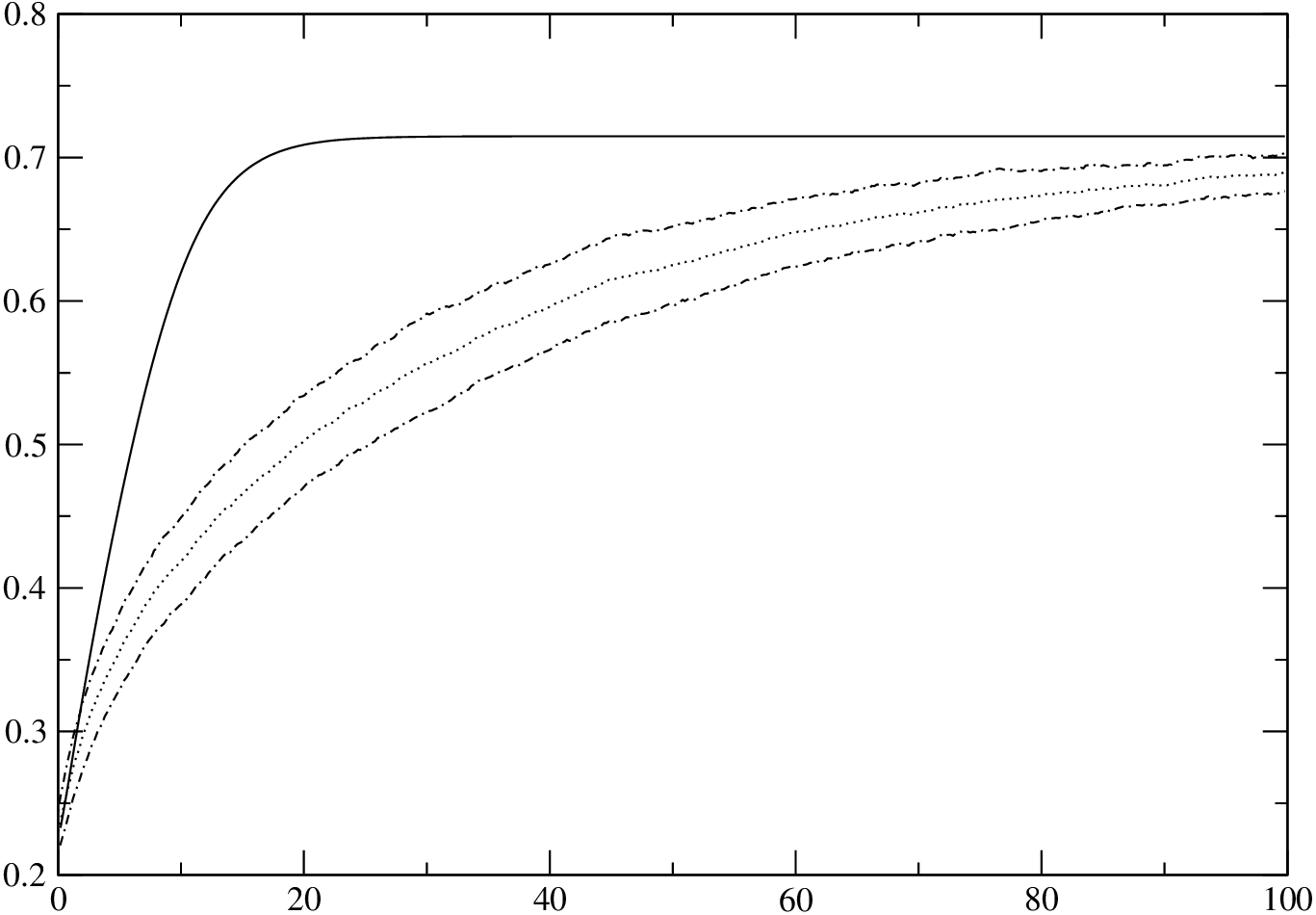}}
\put(-6,90){\epsfysize=60\unitlength\epsfbox{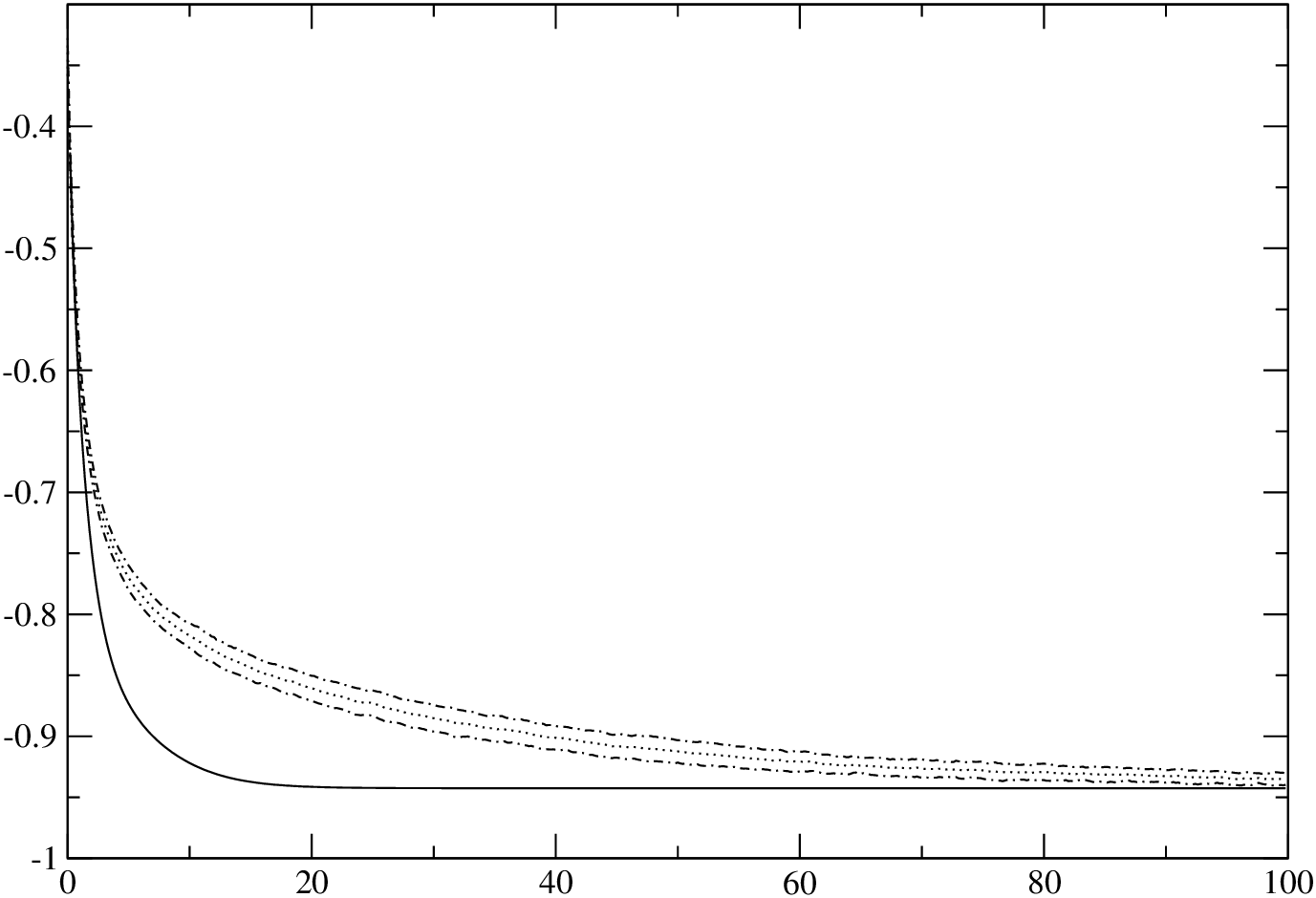}} 
\put(90,90){\epsfysize=60\unitlength\epsfbox{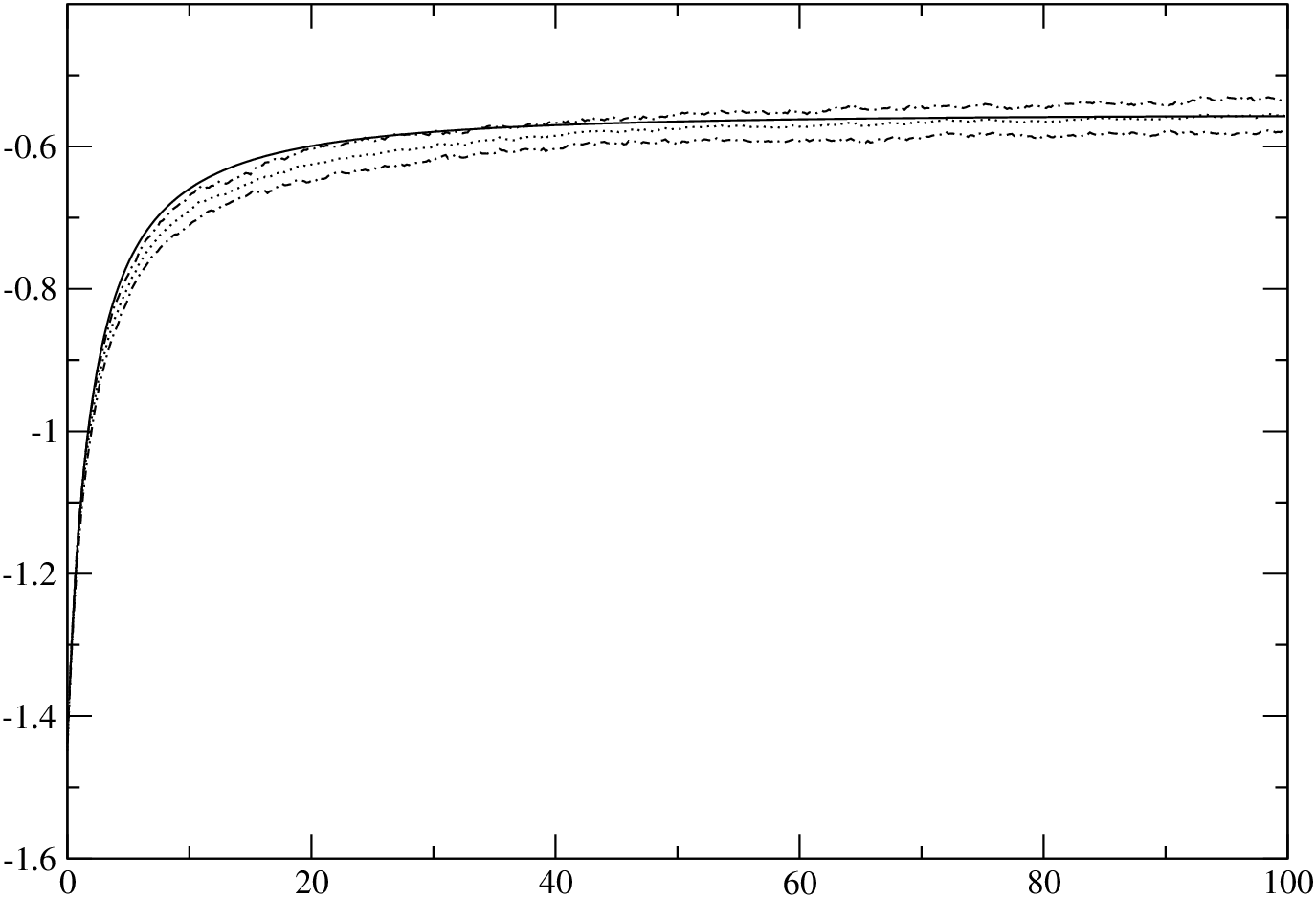}}
\put(90,160){\epsfysize=60\unitlength\epsfbox{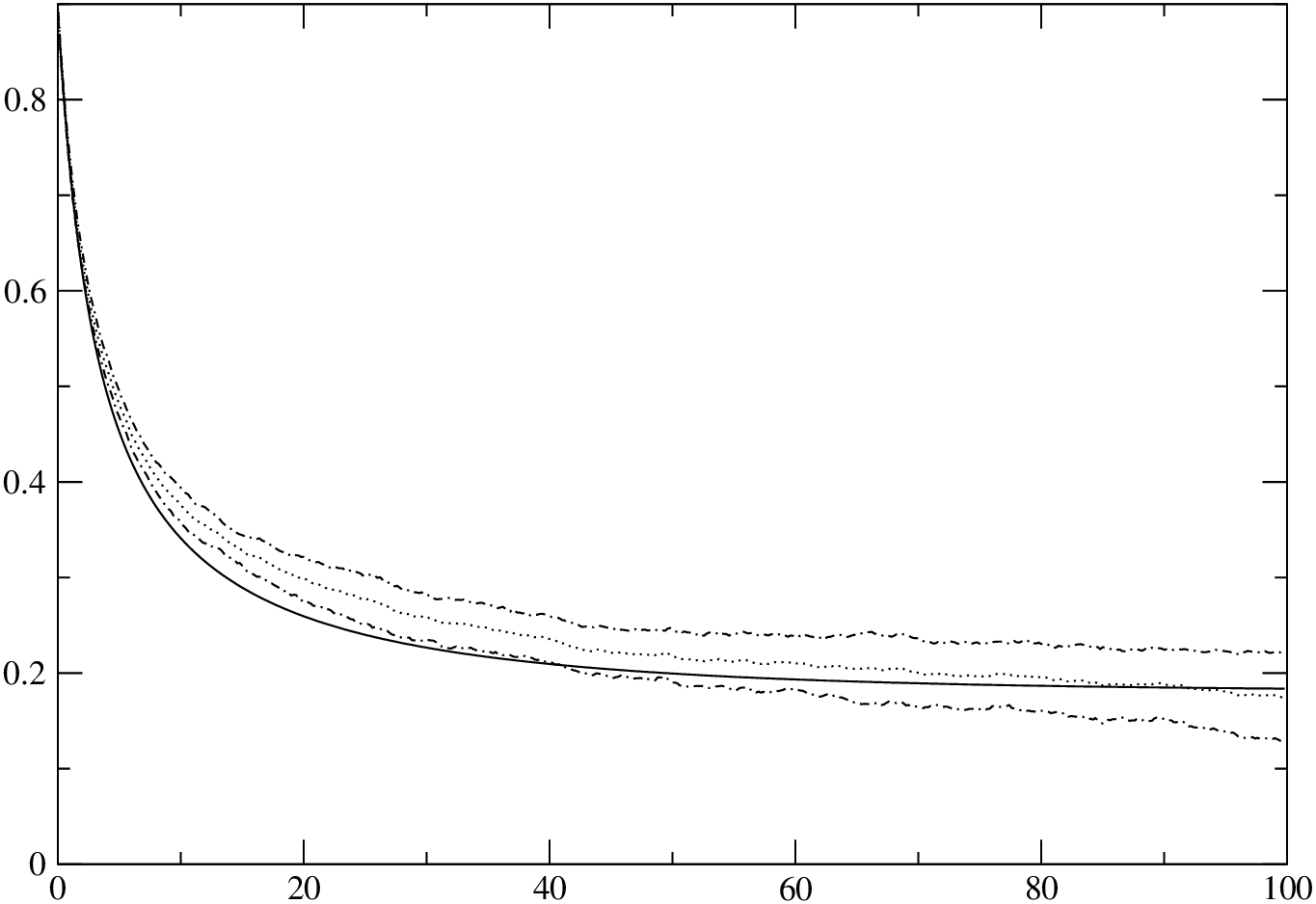}}
\put(43,81){\here{\small updates/spin}}
\put(135,81){\here{\small updates/spin}}
\put(-12,122){\here{$e$}}
\put(-12,192){\here{$m$}}
\put(-12,36){\here{$e$}}
\put(40,2){\here{$m$}}
\put(135,2){\here{$m$}}
\end{picture}
\vspace*{-5mm} \caption{\label{fig:poisson} Evolution of the
magnetization $m$ (top) and the energy $e$ (bottom), for Ising
spins on a Poisson random graph (of average connectivity $c$)
with ferromagnetic bonds $J_{ij}=1$, and with time measured in
units of updates per spin (top 4 graphs) or flow in $m-e$ place
(bottom 2 graphs). Solid lines denote the theoretical
predictions. Dotted lines represent the simulation data (system
size $N=10,\!000$ and averaged over 50 runs), with dot-dashed
lines giving the averages $\pm$ 1 standard deviation. Left
pictures: $c=2$ and $T=0.75$.  Right pictures: $c=3$ and $T=2.8$.
\vspace*{-6mm}}
\end{figure}

In figure \ref{fig:poisson} we examine the order parameter flow in
ferromagnetic random graphs with average connectivity 2 and 3,
respectively. Here each bond is independently defined to be
present ($c_{ij}=1$) or absent ($c_{ij}=0$) with probability
$c/N$, where $c$ is the average connectivity, leading to a graph
with a Poisson degree distribution (or Erd\"{o}s-R\'{e}nyi
graph). In these systems the inhomogeneity of the local
environment of the spins is no longer caused by bond disorder, but
by non-uniform connectivity. The agreement between theory and
simulations in the case $c=2$ is significantly worse than that in
the other examples presented. Here the maximum entropy measure
appears to be a much less accurate approximation of the true
microscopic distribution, which tells us that the system evolves
through statistically a-typical microscopic states, and predicts a
relaxation of the order parameters that is far too quick. This
would appear to be related to the increased heterogeneity
associated with lower temperatures and lower average connectivity.
 However, plotting in the $m-e$ plane we see that the predicted
 direction of the flow is still quite reasonable.

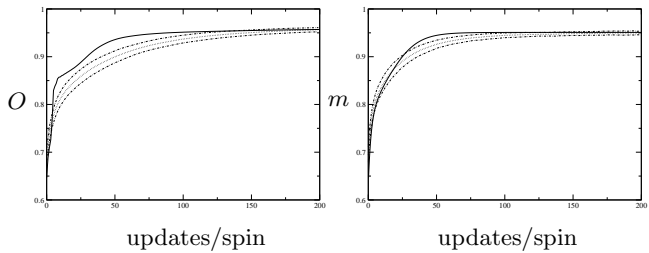
\begin{figure}[t]
\vspace*{5mm} \hspace*{5mm} \setlength{\unitlength}{0.45mm}
\begin{picture}(180,70)
\put(-5,15){\epsfysize=60\unitlength\epsfbox{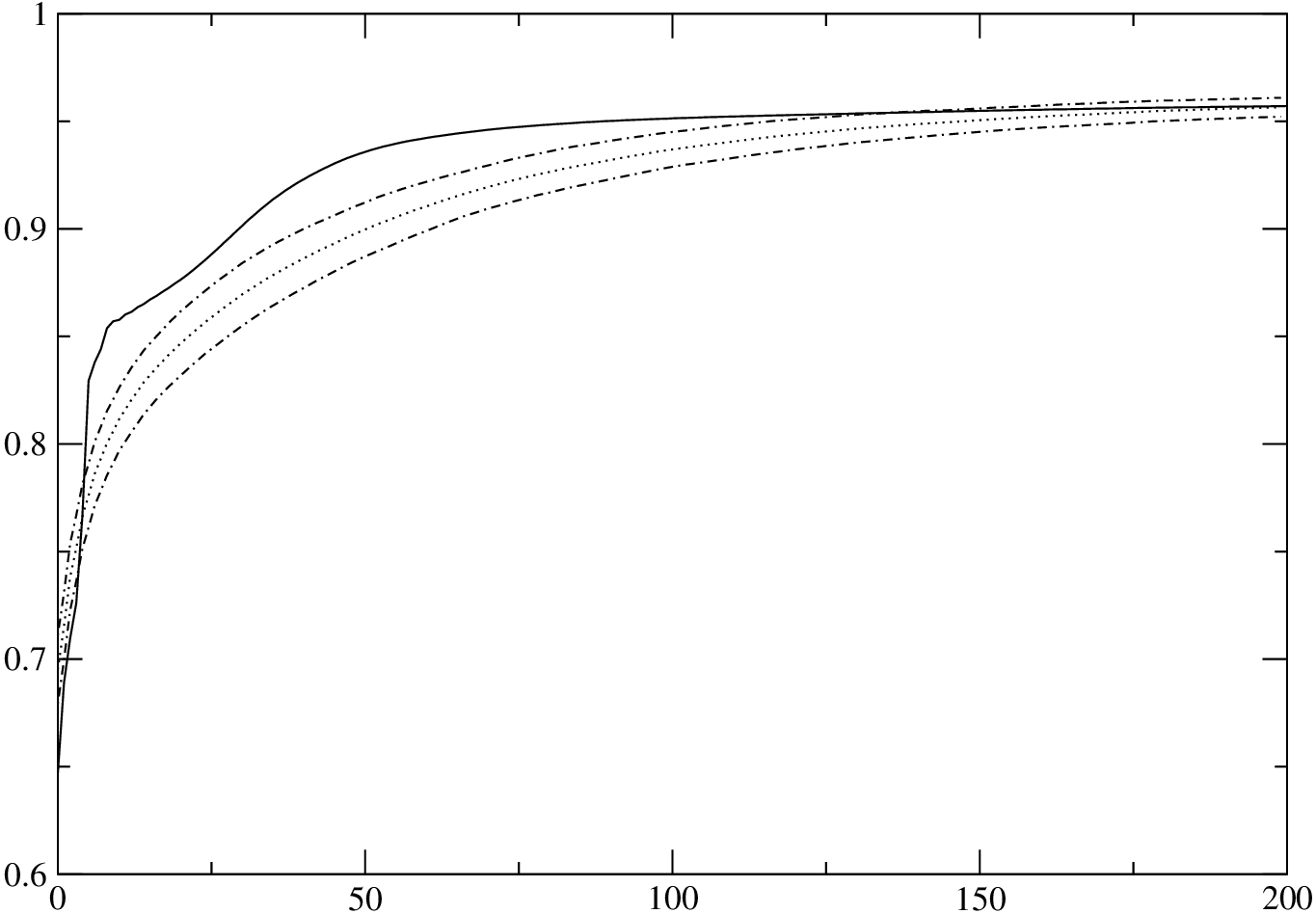}}
\put(90,15){\epsfysize=60\unitlength\epsfbox{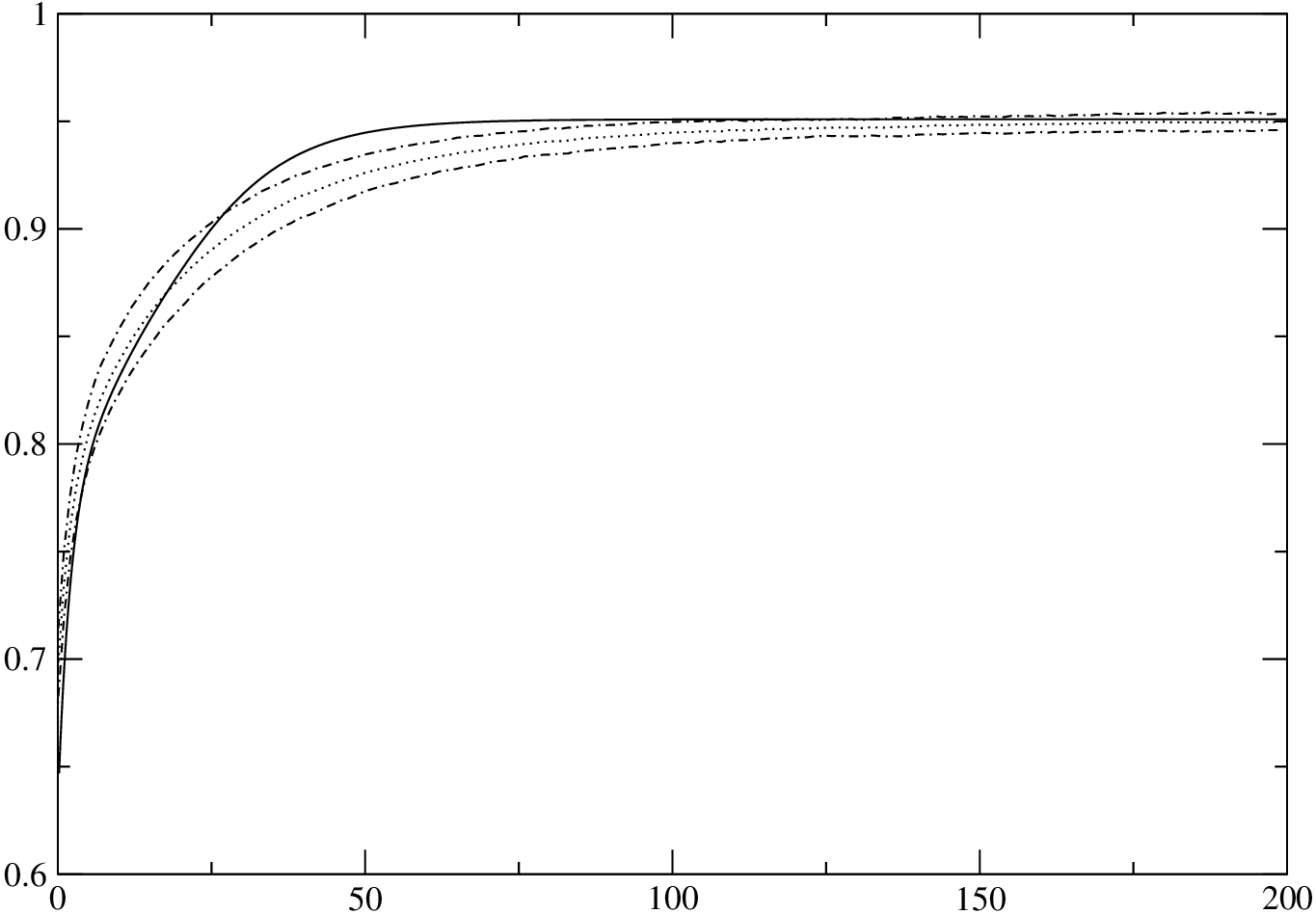}}
\put(43,6){\here{\small updates/spin}}
\put(135,6){\here{\small updates/spin}}
\put(-10,47){\here{$O$}}
\put(85,47){\here{$m$}}
\end{picture}
\vspace*{-5mm} \caption{\label{fig:ecc} Decoding dynamics of the
overlap $O$ (left) and the magnetization $m$ (right),
for a 2-body
interaction and rate $\frac23$ Sourlas error correcting code.
Solid lines denote the theoretical predictions. Dotted lines
represent the simulation data (system size $N=10,\!000$ and
averaged over 50 runs), with dot-dashed lines giving the averages
$\pm$ 1 standard deviation. The temperature is Nishimori's
temperature for the flip probability (error rate) 0.04. 
\vspace*{-6mm}}
\end{figure}

As a final example we turn to the de-coding dynamics of finite
connectivity Sourlas codes \cite{Kabashima1999a,Kanter2000a,Okada2005} with
2-body interactions, which can easily be studied within the
current framework. In particular, in figure \ref{fig:ecc} we
consider the case of an unbiased source broadcasting through a
binary symmetric channel with flip probability 0.04 and rate
$\frac23$ (the channel capacity as given by Shannon's theorem is
0.76...). If a message $(\xi_1,\ldots,\xi_N)$ is sent across this
channel, and our estimator for this message (given the corrupt
channel) is given by $(\hat{\xi}_1,\ldots,\hat{\xi}_N)$, then a
natural performance measure is the overlap between the message
sent and the decoded message, $O = N^{-1}\sum_i \xi_i
\hat{\xi}_i$. We decode at Nishimori's temperature, which is the
temperature maximizing this particular overlap observable
\cite{Rujan1993} (the so-called maximizer of posterior marginals).
Although qualitatively correct, the predicted relaxation of the
order parameters is again too fast compared with the simulation
data.

In this letter  we have presented a relatively simple dynamical
formalism, combining dynamical replica theory with the cavity
method, to be used as a systematic approximation tool with which
to understand the main features of the dynamics of dilute and
disordered spin systems. We regard the wide applicability of the
method as its strength. From the various applications presented
here we see that our approach performs excellently in some cases,
but relaxes too quickly in others, compared with numerical
simulations. This is not unexpected \cite{Coolen1994}. However, as
with the original dynamical replica theory, there is scope for
increasing the order parameter set \cite{Laughton1996,
Semerjian2004b}, which should improve its accuracy systematically,
albeit at a numerical cost.  At present our method requires the
convergence of belief propagation. It would therefore seem that
breaking replica symmetry within this formalism will be
non-trivial to implement, even though theoretically it is a
straightforward generalization. For a single experiment we here
run belief propagation $\mathcal{O}(10^5)$ times; running a finite
temperature 1RSB scheme that many times would be computationally
extremely demanding without further approximations.

We warmly thank M.~Okada, B.~Wemmenhove and T.~Nikoletopoulos for
helpful discussions and comments.

\bibliography{./FOG200501}

\end{document}